\documentclass[%
 reprint,amsmath,amssymb,aps,superscriptaddress]{revtex4-1} 
 
\usepackage[dvipsnames]{xcolor}

\usepackage[T1]{fontenc}
\usepackage[utf8]{inputenc}
\usepackage{graphicx}
\usepackage{dcolumn}
\usepackage{bbm}
\usepackage{ulem}

\begin{document}

\title{Matter-driven change of spacetime topology}

\author{J.~Ambj\o rn}
\email{ambjorn@nbi.dk.}
\affiliation{{\small{The Niels Bohr Institute, Copenhagen University, Blegdamsvej 17, DK-2100 Copenhagen Ø, Denmark.}}}

\author{Z. Drogosz}%
 \email{zbigniew.drogosz@doctoral.uj.edu.pl}
 \author{J. Gizbert-Studnicki }%
 \email{jakub.gizbert-studnicki@uj.edu.pl}
 \author{A. Görlich}%
 \email{andrzej.goerlich@uj.edu.pl}
 \author{J. Jurkiewicz}%
 \email{jerzy.jurkiewicz@uj.edu.pl}
 \author{D. Németh}%
 \email{nemeth.daniel.1992@gmail.com}
\affiliation{%
{\small{Institute of Theoretical Physics, Jagiellonian University, \L ojasiewicza 11, Kraków, PL 30-348, Poland.}}}%

\date{\today}%

\begin{abstract}
Using Monte-Carlo computer simulations, we study the impact of matter fields on the geometry of a typical quantum universe in the CDT model of lattice quantum gravity. The quantum universe has the size of a few Planck lengths and the spatial topology of a three-torus. The matter fields are multicomponent scalar fields taking values in a torus with circumference $\delta$ in each spatial direction, which acts as a new parameter in the CDT model. Changing $\delta$, we observe a phase transition caused by the scalar field. This discovery may have important consequences for quantum universes with nontrivial topology, since the phase transition can change the topology
to a simply connected one.
\end{abstract}

\maketitle

\textit{Introduction} - 
The problem of merging general relativity and quantum mechanics in a theory of quantum gravity has been approached from many directions (string theory \cite{polchinski}, loop quantum gravity \cite{thiemann} and the so-called asymptotic safety program using conventional quantum field theory \cite{asymptoticsafety}, to mention some of the approaches), but no completely satisfactory formulation has yet been found. Difficulties occur already for the pure gravity case, but an additional complication comes from the fact that any realistic theory of quantum gravity should also include coupling to matter fields. The question arises: what type of matter can be included in a particular approach and  what impact does it have on the underlying (quantum) geometric degrees of freedom?  
In this paper we argue that the impact of matter can be quite dramatic even leading to a change of the topology of the Universe.
\vspace{4pt}

\textit{Causal Dynamical Triangulations} - 
Our attempt to examine the above-mentioned question is via a non-perturbative lattice approach to quantum gravity with the name Causal Dynamical Triangulations (CDT)  – see \cite{physrep} for its detailed formulation and \cite{lollreview} for a recent review. It is an approach which lies within the asymptotic safety program and is only using ordinary quantum field theory concepts. In CDT, the (formal) path integral of quantum gravity is lattice-regularized as a sum over 4-dimensional simplicial complexes, called triangulations, which encode geometric degrees of freedom; crucially, they are assumed to be endowed with a causal structure of a globally hyperbolic manifold (i.e., spacetime is foliated into spatial hypersurfaces of fixed and identical topology), which allows a well-defined Wick rotation of the time coordinate. Thus,
\begin{eqnarray}
\label{pathintegral} \mathcal {Z}_{QG} &=& \int\mathcal{D}_{\mathcal{M}_H} [g] \int\mathcal{D}\phi\; e^{i S_{EH}[g]+i S_M[\phi,g]} \to \nonumber \\
&\to& \sum_{T\in\mathcal{T}} \int\mathcal{D}\phi\; e^{- S_R[T]-S_M^{CDT}[\phi, T]}= \mathcal {Z}_{CDT},
\end{eqnarray} 
where  $\mathcal{M}_H$ is a globally hyperbolic Lorentzian manifold, $\mathcal{D}_{\mathcal{M}_H}[ g]$ denotes the integration over geometries, i.e. equivalence classes of metrics $[ g]$ on $\mathcal{M}_H$ with respect to diffeomorphisms, and $\mathcal{T}$ is a suitable set of Wick-rotated (now Euclidean) triangulations. The action $S_R[{T}]$ for a triangulation $T \in\mathcal{T}$ is the Einstein-Hilbert action $S_{EH}$ computed using Regge's method of describing piecewise-linear geometries \cite{regge} and containing the bare couplings related to the cosmological and Newton constants. The second term of the action $S_M^{CDT}[\phi, T]$ is the discrete version of the continuous action $S_M$ for 
matter field(s) $\phi$.
\vspace{4pt}

\textit{Quantum matter fields in CDT} - The simplest quantum matter that can be added to the quantum geometry of CDT is a $d$-component massless scalar field $\phi$. In general, one can assume that the field  $\phi$  has a nontrivial target space, i.e., it is a map  $\mathcal{M}_H(g_{\mu\nu}) \to \mathcal{N}(h_{\alpha\beta})$ between an arbitrary  manifold $\mathcal{M}_H$ (from the path integral (\ref{pathintegral})) with a metric $g_{\mu\nu} $  and  a target space $\mathcal{N}$  with some fixed metric $h_{\alpha\beta}$ and fixed topology. The continuous (Euclidean) action $S_M$ takes the form 
\begin{eqnarray}
S_M[\phi, g]	= \frac{1}{2} \int \mathrm{d}^{4}x \sqrt{g(x)} \; g^{\mu \nu} (x) \; h_{\rho \sigma}(\phi^\gamma(x)) \nonumber \\
\times \partial_\mu \phi^\rho(x) \partial_\nu \phi^\sigma(x).
\label{contmatteraction}
 \end{eqnarray}
Here we  choose the target space  $\cal{N}$ of the scalar field to have either Euclidean $\mathbb{R}^d$ or toroidal $(S^1)^d$ topology, and we fix the flat metric $h_{\rho\sigma}=\delta_{\rho\sigma}$ on $\mathcal{N}$. Consequently, the action (\ref{contmatteraction}) reads
\begin{equation}
S_M[\phi, g]	=\frac{1}{2} \sum_{\sigma=1}^d  \int \mathrm{d}^{4}x \sqrt{g(x)} \;  \partial^\nu \phi^\sigma(x) \partial_\nu \phi^\sigma(x),
\label{matteraction}
\end{equation}
and the various components decouple for different $\sigma$ because the target space metric is diagonal. For a particular sample  geometry $[ g]$, quantum fluctuations of $\phi^\sigma$ will occur around a semi-classical solution $\bar \phi^\sigma$ satisfying the Laplace equation
\begin{eqnarray}
\Delta_x \bar \phi^\sigma(x) = 0,\quad 
\Delta_x = \frac{1}{\sqrt{g(x)}}\partial_\mu \sqrt{g(x)}g^{\mu\nu}(x)\partial_\nu.
\label{laplace_eq1}
 \end{eqnarray}
Let us  {now} start with a simple case where the target space of the scalar field is $\mathcal{N} = \mathbb{R}^d$. In CDT we consider the scalar field as located at the centers of  equilateral simplices, and thus the discrete counterpart of the action~(\ref{matteraction})  {takes a very simple form}
\begin{equation}
S_M^{CDT}[\phi, T] =\frac{1}{2} \sum_{\sigma=1}^d \sum_{i \leftrightarrow j}(\phi^\sigma_i-\phi^\sigma_j)^2=  \sum_{\sigma,i,j}  \phi_i^\sigma \mathbf{L}_{ij}(T) \phi_j^\sigma.
\label{action}
\end{equation}
The sum  $ \sum_{i \leftrightarrow j}$ is over the five pairs of neighboring four-simplices of each simlex $i$ in the triangulation $T$ of the manifold $ \mathcal{M}_H(g_{\mu\nu})$ and $\mathbf{L}=5 \mathbbm{1} - \mathbf{A}$ is the Laplacian matrix, where $\mathbf{A}_{ij}$ is the adjacency matrix with entries of value 1 if simplices $i$ and $j$ are neighbors and 0 otherwise.  The discrete version of the Laplace eq. (\ref{laplace_eq1}) for each component of the classical scalar field is then 
\begin{equation}\label{laplace_eq2} 
    \mathbf{L}\bar\phi^\sigma =  0,
\end{equation}
which is solved by $\bar \phi = \mathrm{const}.$ (the Laplacian zero mode) for any compact simplicial manifold $T$.
After the decomposition of the field
\begin{equation}\label{decompose}
\phi^\sigma = \bar \phi^\sigma+ \xi^\sigma
\end{equation}
into the classical part $ \bar \phi^\sigma$ and the quantum part $ \xi^\sigma$ and an application of (\ref{laplace_eq2}), the contribution from the classical field vanishes, leaving
\begin{equation}\label{action2}
S_M^{CDT}[\phi, T] = \frac{1}{2} \sum_{\sigma=1}^d \sum_{i \leftrightarrow j}(\xi^\sigma_i-\xi^\sigma_j)^2
= \sum_{\sigma,i,j}  \xi_i^\sigma \mathbf{L}_{ij}(T) \xi^\sigma_j.
\end{equation}
The Gaussian form of the matter action (\ref{action2}) means that, in principle, the field can be integrated out,
contributing to the geometric action $S_R[T] \to S_R[T]+ S_{M}^{eff}[T]$  with a term 
\begin{equation}
\label{det}
 S_{M}^{eff}[T] = \frac{d}{2}\log \det({\mathbf L'}( T)),
 \end{equation}
where ${\mathbf L'}(T)$ is the Laplacian matrix $\mathbf L( T)$ in the subspace orthogonal to the constant zero mode of $\mathbf{L}$. 
The dependence of  eq.~(\ref{det}) on {the} geometry rests in the dependence of ${\mathbf L'}( T)$ on the adjacency matrix $\mathbf A$ defined for a given triangulation $T$. 
Using numerical Monte-Carlo simulations we checked that 
the dependence of the determinant $S_{M}^{eff}[T]$ on $T$ is weak and, in practice, we can treat it as a constant. 

\vspace{4pt}

\textit{Quantum scalar fields with values on $(S^1)^d$ } - 
The new aspect studied here is based on two major generalizations of the CDT model:

(1)
We choose the spatial topology of the time-foliation leaves to be $(S^1)^3$, and for technical reasons we assume the time boundary conditions to be periodic as well. Thus, each triangulation has the toroidal topology $(S^1)^4$ and can equivalently be represented as an elementary cell periodically repeated in four dimensions. There is a lot of freedom in the selection of the elementary cell; one way to determine it is to choose four independent non-contractible three-dimensional boundaries delimiting it. The boundaries are connected sets of three-dimensional faces, each shared by two four-simplices located in different copies of the elementary cell. We assume each boundary to be oriented, and we encode the information about the position of the four boundaries (labeled by $\sigma=1,2,3,4$) in a triangulation $T$ within four matrices $\mathbf{B}^\sigma$, whose elements are
\begin{equation}
\label{adjacency}
\mathbf{B}_{ij}^\sigma =
\begin{cases}
	\pm 1 & \textrm{if the face shared by simplices $i$ and $j$}\\
	 & \textrm{exists and belongs to the boundary}\\
	0 & \textrm{otherwise}.
\end{cases}
\end{equation}

The number of directed boundary faces of a simplex $i$ is given by $b_i^\sigma=\sum_j \mathbf{B}_{ij}^\sigma$, and the boundary three-volume is $V^\sigma=\frac{1}{2}\sum_{ij}\left(\mathbf{B}^\sigma_{ij}\right)^2$. Despite being fictitious constructs having no impact on the physics, the boundaries can be used to define a coordinate system, as described in \cite{coordinates, class-letter}.

\begin{figure}
\centering
\includegraphics[scale=0.13]{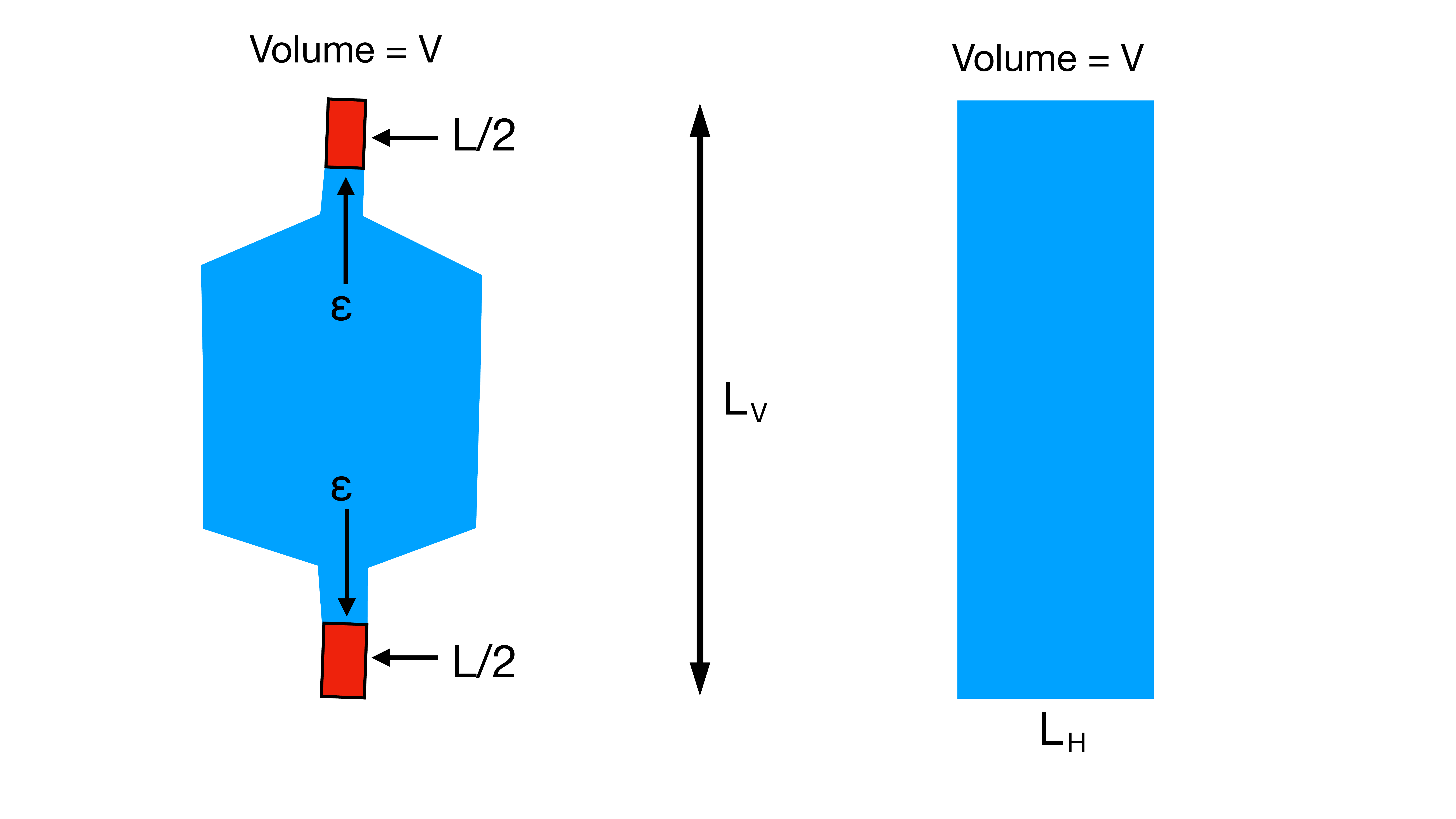}
\caption{Left: a torus (opposite sides identified) with a pinch. The region in red is the region where $\phi$ changes from 0 to $\delta$.
 In the blue part it stays constant.
$\phi$ is constant in the horizontal direction. 
Right:  a torus where $\phi$ is constant in the horizontal direction and uniformly increases from 0 to $\delta$ from bottom to top. }
\label{fig-pinch}
\end{figure}

(2) The $d$-component scalar field $\phi$ is assumed to take values on a symmetric torus $\mathcal{N}=(S^1)^d$ with  circumference $\delta$ in each direction. We require that each component of the field $\phi^\sigma \in S^1$ winds around the circle  once as we go around any non-contractible loop in $T$ that crosses a boundary  in direction $\sigma$. This requirement completely changes the dynamics of the interaction between geometry and the scalar field. For the scalar field taking values in $\mathbb{R}^d$, the classical solution is constant and does not contribute to the matter action, which depends therefore only on quantum fluctuations. For $\phi^\sigma \in S^1$, however, the constant solution is not allowed, since it has winding number zero; as we will show below, the new non-trivial classical solution does contribute to the effective matter action. The winding condition can technically be obtained by considering a field $\phi^\sigma \in \mathbb{R}$ that jumps by $ \delta\cdot\mathbf{B}^\sigma_{ij}$ on any face shared by simplices $i$ and $j$ 
and identifying
\begin{equation}\label{identify}
 \phi_i^\sigma \equiv \phi_i^\sigma+n \cdot \delta, \quad n\in \mathbb{Z}.
 \end{equation}
The (discrete) matter action becomes
\begin{eqnarray}
 S_M^{CDT}[\phi,T] = \frac{1}{2} \sum_{\sigma=1}^d \sum_{i \leftrightarrow j}(\phi^\sigma_i-\phi^\sigma_j-\delta \cdot \mathbf{B}^\sigma_{ij})^2 \label{actionb}
\end{eqnarray}
leading to the following equation for the classical field $\bar\phi^\sigma$
\begin{equation}\label{laplace_eq3} 
    \mathbf{L}\bar\phi^\sigma =  \delta \cdot b^\sigma,
\end{equation}
 which now acquires a boundary term and thus admits nontrivial solutions for $\bar\phi^\sigma$. Note that the action (\ref{actionb}) is invariant under a  local shift of the $\sigma$-boundary 
 with a simultaneous change of the scalar field value  $\phi_i^\sigma \to \phi_i^\sigma\pm \delta$. 
Thus  the choice of a specific boundary  does not influence the path integral (\ref{pathintegral}) in any way. Decomposition (\ref{decompose}) of the  field into the classical and the quantum part yields
\begin{eqnarray}
 S_M^{CDT}[\phi,T] 
 = \sum_{\sigma,i,j}  \xi_i^\sigma \mathbf{L}_{ij}(T) \xi_j^\sigma + S_M^{CDT}[\bar \phi,T]. 
 \label{actionb2}
\end{eqnarray}
Since $\phi^\sigma$ and  $\bar \phi^\sigma$ have winding number one, the fluctuation field $\xi^\sigma$ is a scalar field with winding number zero, i.e., an ordinary scalar field taking values in $\mathbb{R}$. The action (\ref{actionb2}) is then  again Gaussian and can be integrated out, 
now leading to $S_R[T]\to S_R[T]+S_M^{eff}[T] + S_M^{CDT}[\bar\phi,T]$, where the determinant $S_M^{eff}[T]$ is the same as in eq.~(\ref{det}). Thus the only difference between the impact of the scalar field in $(S^1)^d$ and that of the ordinary field in $\mathbb{R}^d$ is the dependence of the effective matter action on the non-trivial classical solution $\bar\phi$; note that the size of the jump $\delta$ fixes the scale of the classical field. For $\delta=0$ one recovers the $\mathbb{R}^d$ case where $S_M^{CDT}[\bar\phi,T]$ is zero (the absolute minimum), as $\bar\phi=\mathrm{const}.$ for any quantum geometry $T$. For $\delta>0$ the constant solution is not allowed, and the action $S_M^{CDT}[\bar\phi,T]$ depends on the specific geometry $T$.  By adjusting the geometry in a rather drastic way, one is still able to reduce the matter action almost to zero. This is illustrated in Fig.\ \ref{fig-pinch} in the simple case of a two-dimensional torus with a one-dimensional field $\phi$ changing in the vertical direction, but the argument is clearly valid in higher dimensions, and in fact it only depends on one direction being periodic. The topology in the ``transverse'' directions can be anything.  
On the left plot we have a torus with volume $V$ and vertical length $L_V$,
which is pinched to a cylinder of circumference $\varepsilon$ and length $L$.
The total matter action of the field configuration alluded to on the 
plot is  
\begin{equation}\label{jan50}
S_{M}^{CDT}[\phi, T_L] = \Big( \frac{\delta}{L}\Big)^2 L \, \varepsilon = \delta^2 \frac{\varepsilon}{L},
\end{equation}
and the minimal action for a classical field configuration $S_{M}^{CDT}[\bar\phi,T_L]$ for this geometry is even lower. This can clearly be made arbitrarily small when $\varepsilon \to 0$, and this is even more true in higher dimensions. On the right plot we also have a torus with volume $V$ and vertical length $L_V$. For this geometry, the action is minimal for a field changing uniformly from 0 to $\delta$, when we move from bottom to top, thus obtaining an action 
\begin{equation}\label{jan51}
S_{M}^{CDT}[\bar\phi,T_R] = \Big( \frac{\delta}{L_V}\Big)^2 L_V L_H  = \delta^2 
\frac{V}{L_V^2},~ V= L_H L_V,
\end{equation}
which is bounded from below when $V$ and $L_V$ are fixed. Let us discuss the consequence of this in the full quantum theory defined by the path integral (\ref{pathintegral}). The classical action $S_M^{CDT}[\bar\phi,T]$ depends in a crucial way on the triangulation $T$. The triangulations that are pinched as shown in Fig.\ \ref{fig-pinch} will have the smallest matter action, but the geometric Einstein-Hilbert part of the action will be larger for such pinched configurations than for ``regular'' triangulations. A simple minisuperspace model, like the Hartle-Hawking model \cite{hh}, suggests that for small jumps $\delta$ the geometric part of the action dominates and the generic configurations in the path integral are quite similar to the ones which dominate when no matter field with a jump is present. However, for large $\delta$ the total action will be the lowest for pinched configurations and the system will instead fluctuate around pinched configurations. Thus, the system might undergo a phase transition as a function of the jump magnitude $\delta$.

\vspace{4pt}


\textit{Results for scalar fields winding around spatial directions} -
Below, using numerical Monte Carlo simulations, we study a CDT model with a $d=3$ component massless scalar field taking values in a symmetric torus $\mathcal{N}=(S^1)^3$ with circumference $\delta$, i.e., $\phi^\sigma$ jumps by $\pm \delta$ when crossing a 3-dimensional boundary orthogonal to one of  three independent non-contractible loops  winding around the toroidal spatial direction $\sigma=x,y,z$ in a triangulation $T$. This type of matter system was earlier introduced to define a semiclassical coordinate system for a given CDT triangulation \cite{class-letter}.
Now we want to make the scalar field a dynamical (quantum) object and let it evolve together with the geometry. Then, for a given geometric configuration $T$, one  can simply compute the expectation value of the field $\langle \phi \rangle=\bar\phi$ by solving the Laplace equation (\ref{laplace_eq3}) and  use it as spatial coordinates. The analyzed systems were all in the same point in the CDT parameter space inside the so-called semiclassical (or de-Sitter) phase \cite{semiclassical,c-phase1,c-phase2}, and the only variable parameter was the  jump magnitude $\delta$. For each  analyzed value of $\delta$  we pick a generic quantum geometry (a triangulation $T$) appearing in the path integral (\ref{pathintegral}), and we use the methodology introduced in \cite{class-letter} to assign a unique set of spatial coordinates  for each simplex $i$ defined by the  classical solution of the scalar field $(\bar\phi^x_i,\bar\phi^y_i,\bar\phi^z_i)$ computed for that geometric configuration. Note that in CDT one has the time coordinate ``for free'' as the classical solution  of the field $\bar\phi_i^t$ can be computed using the imposed proper-time foliation. In order to visualize  changes in a typical quantum geometry triggered by the increasing jump magnitude, we measure the four-volume density distribution $N(\bar\phi)$, i.e., the number of simplices contained in hypercubic blocks with sizes $(\Delta\bar\phi^x_i,\Delta\bar\phi^y_i,\Delta\bar\phi^z_i,\Delta\bar\phi^t_i)$, which is equivalent to measuring the integrated $\sqrt{g(\bar\phi)}$:
\begin{eqnarray}
\Delta N(\bar\phi) = \sqrt{g(\bar\phi)} \prod_\sigma \Delta \bar\phi^\sigma = N(\bar\phi) \prod_\sigma \Delta \bar\phi^\sigma.
\label{hypers_vol_eq_psi}
\end{eqnarray}

\begin{figure}[t]
\includegraphics[width=0.4\textwidth]{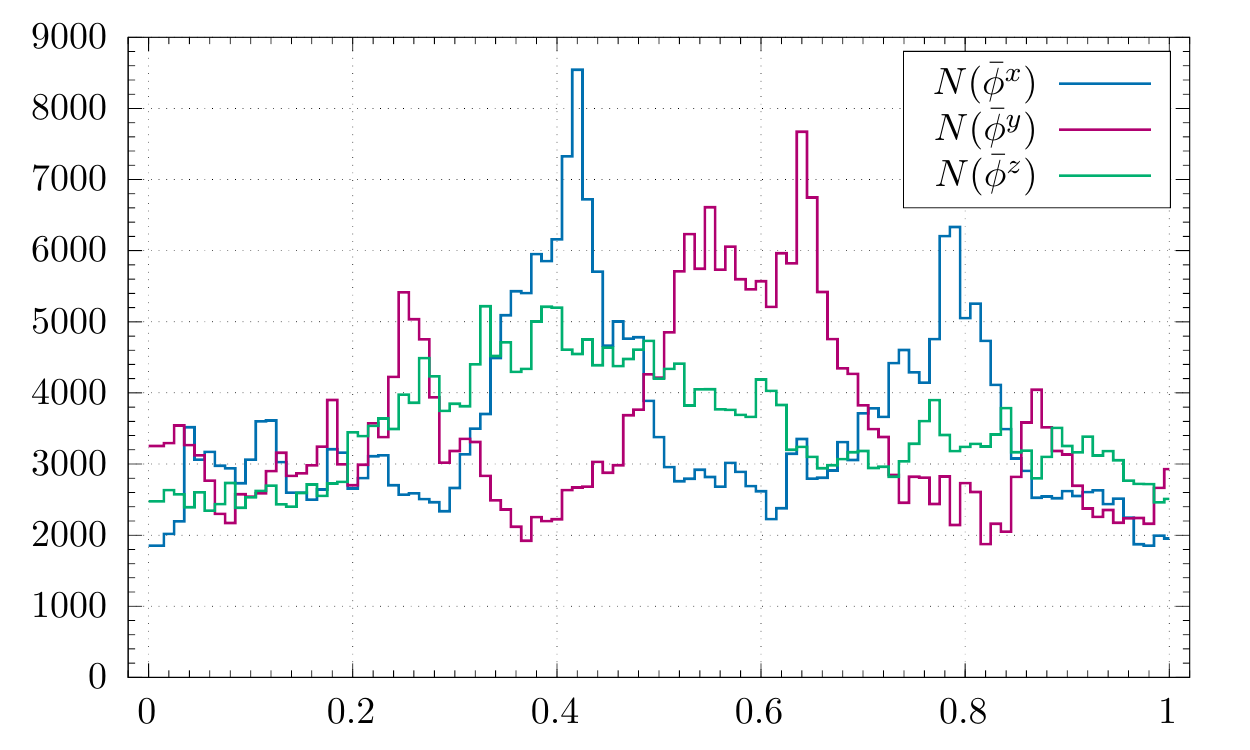} 
\caption{The projection of four-volume, as defined by (\ref{hypers_vol_eq_psi}),  on one spatial direction ($x$, $y$ or $z$) for a typical  CDT configuration  with  small jump magnitude ($\delta=0.1$).
{The horizontal axis is $\bar{\phi}/\delta$.}}\label{1dsmall}
\end{figure}

\begin{figure}[t]
\includegraphics[width=0.4\textwidth]{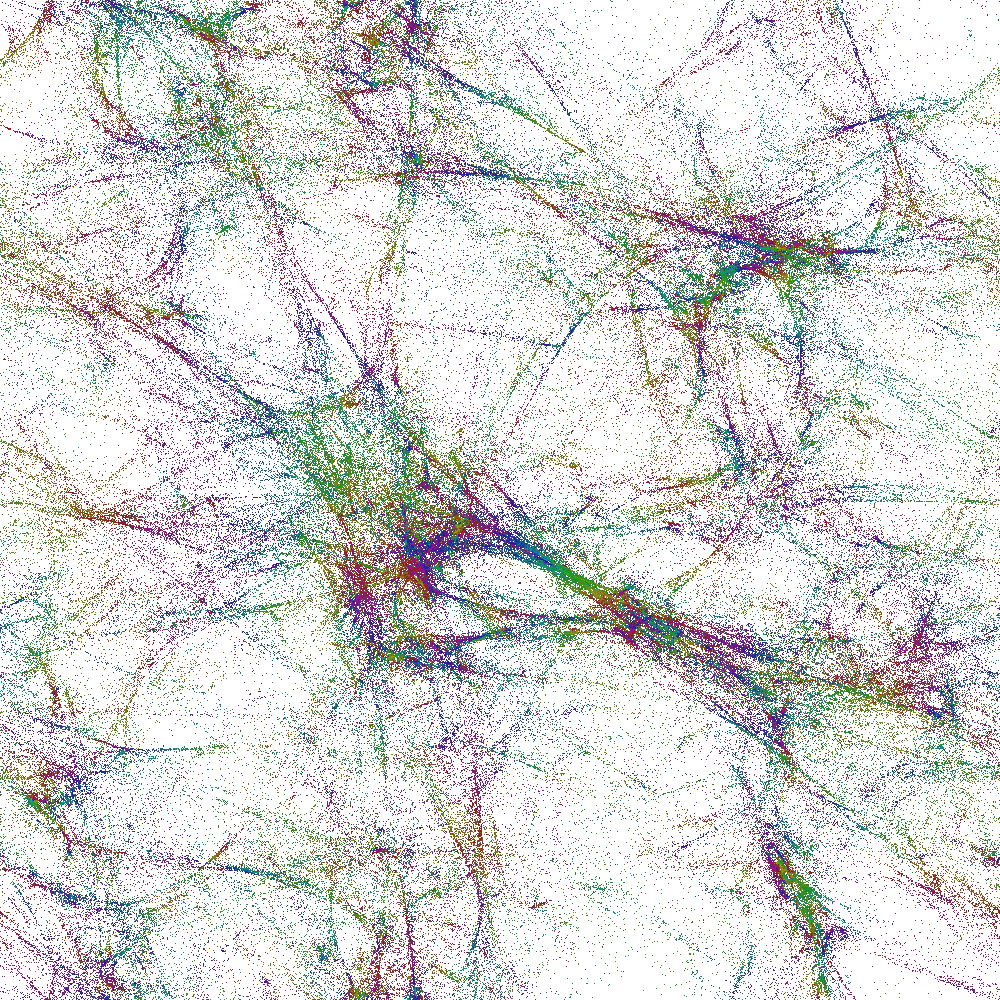} 
\caption{The projection of four-volume,  as defined by (\ref{hypers_vol_eq_psi}), on the $xy$-plane for a typical CDT configuration with  small jump magnitude ($\delta=0.1$).
Different colors correspond to different times $t$ of the original (lattice) time-foliation.}\label{xysmall}
\end{figure}

\begin{figure}[t]
\includegraphics[width=0.4\textwidth]{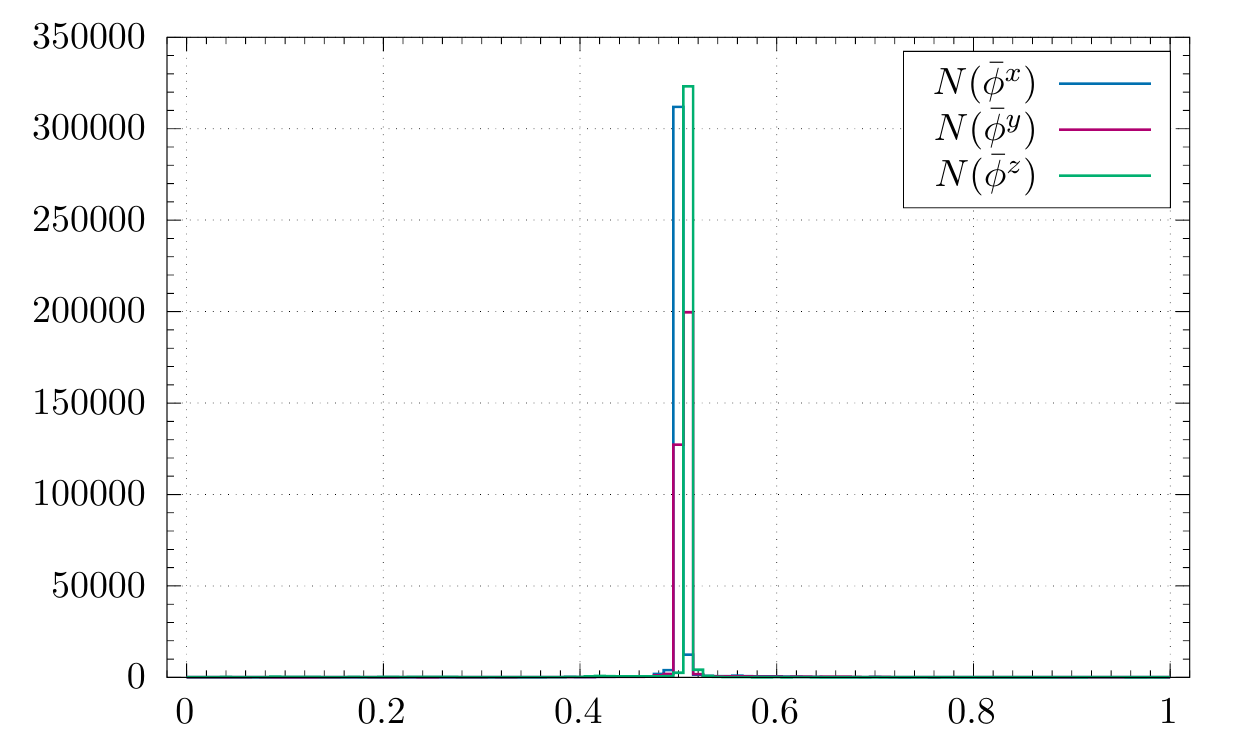}
\caption{The projection of four-volume, as defined by (\ref{hypers_vol_eq_psi}),  on one spatial direction ($x$, $y$ or $z$) for a typical  CDT configuration  with large jump magnitude ($\delta=1.0$). {The horizontal axis is $\bar{\phi}/\delta$}.}
\label{1dlarge}
\end{figure}

\begin{figure}[t]
\includegraphics[width=0.4\textwidth]{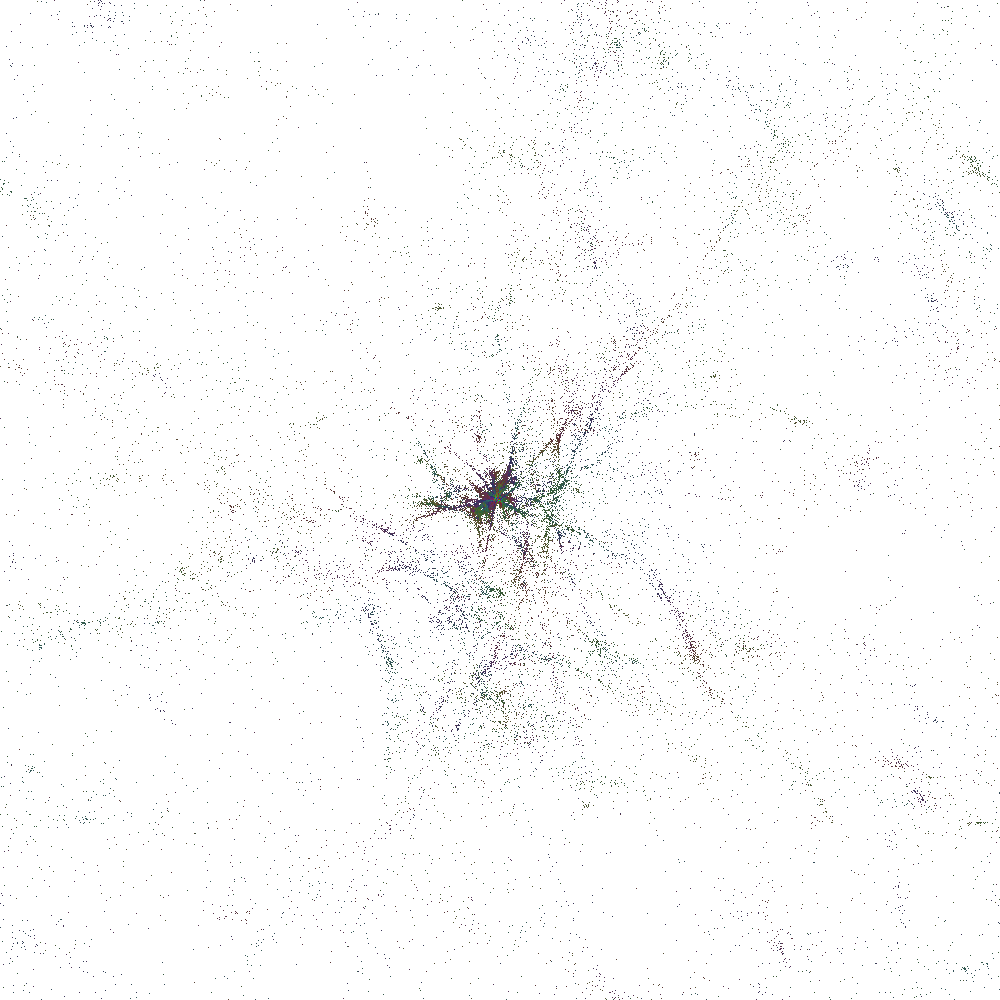} 
\caption{The projection of four-volume, as defined by (\ref{hypers_vol_eq_psi}), on the $xy$-plane for a typical  CDT configuration  with  large jump magnitude ($\delta=1.0$).
Different colors correspond to different times $t$ of the original (lattice) time-foliation. 
}\label{xylarge}
\end{figure}

In Figs. \ref{1dsmall} and \ref{1dlarge} we plot projections of the volume density  distribution $N(\bar\phi)$ in a typical toroidal CDT configuration on one spatial direction ($x$, $y$ or $z$), and in Figs. \ref{xysmall} and \ref{xylarge} the projections on two-dimensional ($x-y$ directions) parameter subspace, integrating over the remaining directions. Figs. \ref{1dsmall} and \ref{xysmall} are for a small ($\delta=0.1$), while Figs. \ref{1dlarge} and \ref{xylarge} for a relatively large ($\delta=1.0$) jump magnitude, respectively. One  clearly observes 
that the change of the jump magnitude causes a substantial change in a typical CDT geometry. For a small jump ($\delta=0.1$) one observes a geometry which resembles the pure gravity case, see \cite{class-letter}. For a large jump ($\delta=1.0$), in line with expectations, one observes that the geometry is ``pinched'' in all spatial directions (which manifests itself as the small-volume region in Fig. \ref{1dlarge} and the low-density region in Fig. \ref{xylarge}).
\vspace{4pt}

\textit{Discussion} - 
We have shown that if spacetime is globally hyperbolic and has the toroidal spatial topology, i.e., has three nonequivalent non-contractible loops in the spatial directions, then the three-component scalar field with matching topological boundary conditions imposed can have a dramatic effect on the geometries that dominate the CDT path integral. If the spatial topology is simply connected, this effect is absent. This new kind  of coupling between the topology of the matter fields and the topology of spacetime is likely to result in a phase transition for sufficiently strong coupling (sufficiently large $\delta$ in our model), a transition where the path integral will be dominated by spatial geometries with pinched regions fluctuating close to zero sizes (but still connected due to topological restrictions imposed by our model). This is schematically shown in Fig.\ \ref{fig-pinch} and for actual configurations in the path integral in Figs.~\ref{1dlarge} and \ref{xylarge}, 
using as coordinates in the non-contractible ``directions'' the classical scalar fields with non-trivial boundary conditions in these directions, and what is also visible using other coordinate systems, e.g., the ones introduced in \cite{coordinates}. Extrapolating this result to  large volume limit we get a picture with a small toroidal part of cut-off size and the dominating geometry with an (almost) spherical topology. Concluding, the effect of scalar fields can be much more drastic than previously appreciated, with possible implications for cosmological model building and even in other areas of physics related to phase transitions of topological nature.
\vspace{4pt}

\begin{acknowledgments}
Z.D. acknowledges support from the National Science Centre, Poland, grant 2019/32/T/ST2/00390. J.G.-S. acknowledges support of the grant UMO-2016/23/ST2/00289 from the National Science Centre Poland. A.G. acknowledges support by the National Science Centre, Poland, under grant no. 2015/17/D/ST2/03479. J.J. acknowledges support from the National Science Centre, Poland, grant  2019/33/B/ST2/00589. D.N. acknowledges support from National Science Centre, Poland with grant no. 2019/32/T/ST2/00389. 
\end{acknowledgments}


\end{document}